\documentclass[]{article}

\usepackage{graphicx}
\usepackage{float}
\usepackage{url}
\usepackage{authblk}
\title{\textbf{Correcting for Leakage Energy in the SiD Silicon-Tungsten ECal} \\ \vspace{0.25in} \large \textit{Talk presented at the International Workshop on Future Linear Colliders (LCWS2019), Sendai, Japan, 28 October-1 November, 2019. C19-10-28.}}
\author[]{L. Braun, D. Austin, J. Barkeloo, J. Brau, C.T. Potter}
\affil[]{Physics Department, University of Oregon }

\date{\today}

\begin{document}

\twocolumn

\maketitle

\begin{abstract}
A dominant contribution to ECal resolution at high energy (eg. 100 GeV) comes from leakage beyond the containment of the calorimeter. We have studied the leakage energy for the SiD silicon-tungsten ECal and developed a neural network algorithm for estimating the leakage energy and correcting the energy measurement.  The SiD TDR design calls for 20 thin 2.5 mm tungsten layers followed by 10 thick 5.0 mm tungsten layers. We have investigated the impact on the leakage energy of a reduced number of layers, and the ability of an optimized neutral network analysis to correct for the leakage with a reduced number of layers, and reduced material thickness. Reducing layer numbers is motivated by cost containment.
\end{abstract}

\section{Introduction}

The International Linear Collider (ILC) has emerged as the most technically mature proposal for the next generation $e^+ e^-$ collider. The primary motivation for building the ILC is precision study of the Higgs boson, the particle recently discovered in 2012 at the Large Hadron Collider (LHC) \cite{Aad:2012tfa,Chatrchyan:2012ufa} but first hypothesized in 1964 as the particle associated with the field responsible for spontaneous electroweak gauge symmetry breaking \cite{Higgs:1964pj,Englert:1964et}.

The Silicon Detector (SiD) is one of two technically mature ILC detector proposals detailed in the SiD Letter of Intent (LoI) \cite{Aihara:2009ad} and ILC Technical Design Report (TDR) \cite{Behnke:2013xla,Baer:2013cma,Phinney:2007gp,Behnke:2013lya}. It features a 5T solenoid with sufficiently precise calorimetry to yield the high jet energy resolution necessary for measuring Higgs boson branching ratios, including those to exotic or invisible final states, to unprecedented precision.

In the LoI a cost optimization of the hadronic calorimeter (HCal) radius and depth, together with the solenoid field strength, was performed, leading to the nominal design detailed in both the LoI and the TDR. This optimization left the electromagnetic calorimeter (ECal) as one of the most expensive components of SiD for both material and labor. 

The energy resolution of the ECal is also critically important for Higgs boson branching ratios to final states with electrons and photons. The material costs of the ECal are dominated by the cost of high grade Silicon and Tungsten. In this work we determine whether a reduced cost design of the SiD ECal, with fewer Silicon and Tungsten layers, can maintain the necessary high performance of the nominal design by recovering ECal energy leakage. 

\section{SiD ECal}

\begin{figure*}[t]
\hspace{0.5in}\includegraphics[angle=90,width=0.35\textwidth]{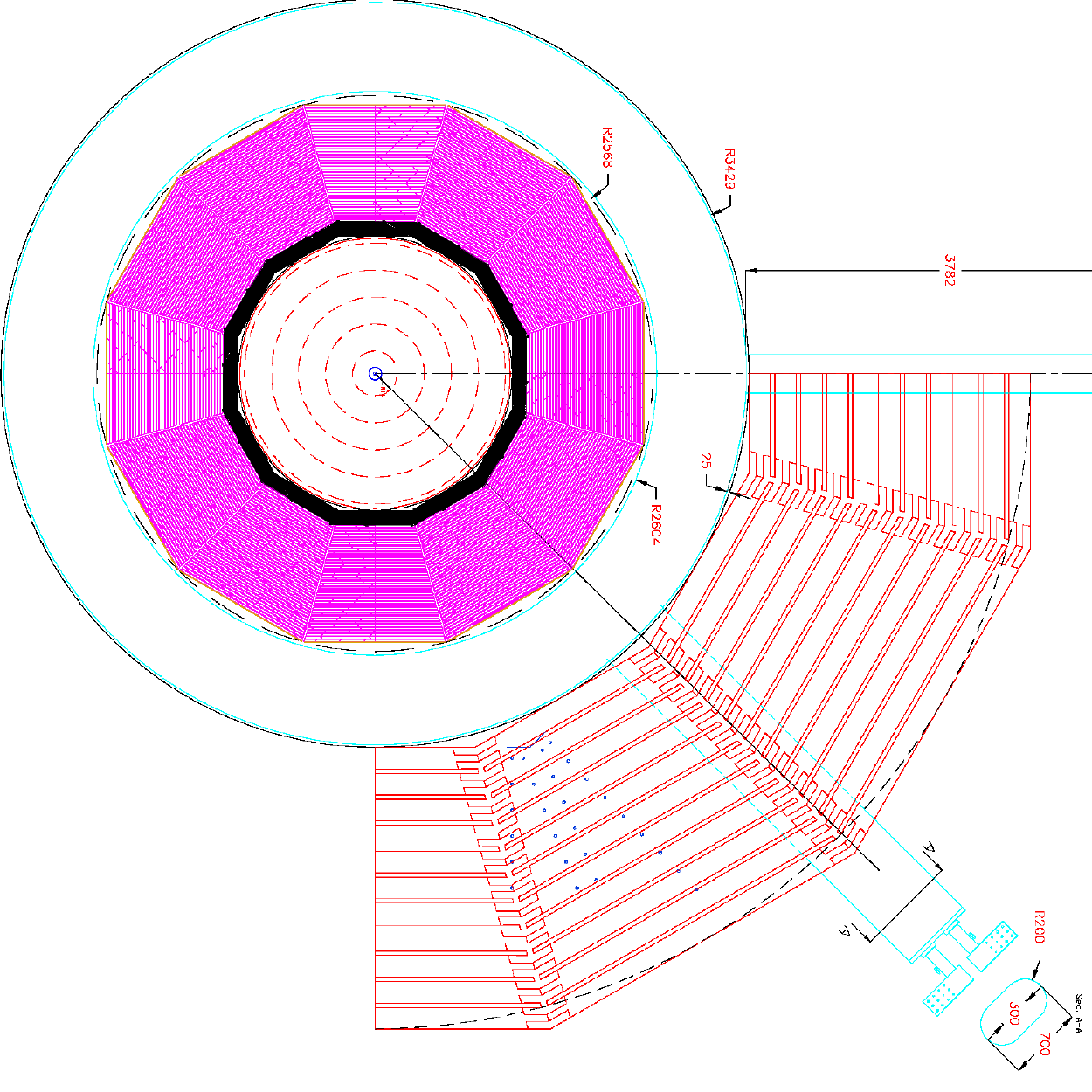}\hspace{0.5in}
\includegraphics[angle=90,width=0.5\textwidth]{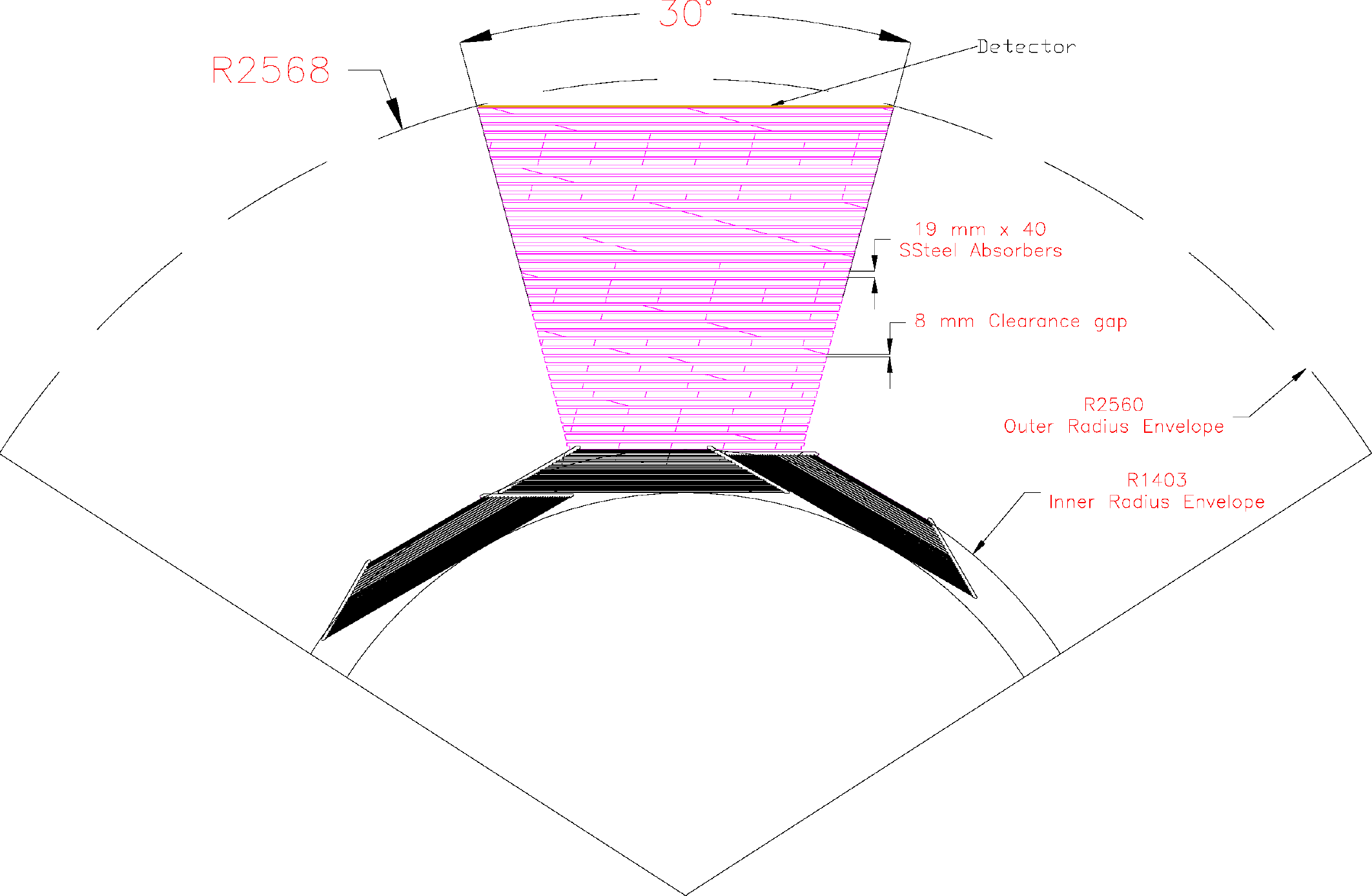}
\caption{At left, barrel view of SiD. At right, barrel view of the nominal SiD ECal (black) and HCal (magenta) designs. Note the ECal module overlap regions recur every 30$^{\circ}$ in $\phi$.}
\label{fig:ecal}
\end{figure*}

In the nominal SiD ECal design, absorbing Tungsten layers alternate with sensitive Silicon layers, with 20 thin (2.5mm) Tungsten layers followed by 10 thick (5.0mm) layers for a total of $26X_0$ in both barrel and endcaps. In the barrel, twelve trapezoidal modules are designed to overlap in order to provide mechanical stability and cover projective gaps. See Figure \ref{fig:ecal} for a technical drawing of the nominal ECal design.

The nominal SiD ECal energy resolution is given by

\begin{eqnarray}
\frac{\Delta E}{E} & = & 0.01 \oplus \frac{0.17}{\sqrt{E}}
\end{eqnarray}

\noindent However, some electromagnetic showers develop late and leak energy into the HCal, and this problem is exacerbated for designs with fewer layers. Thus ECal resolution is limited by energy leakage.

In this study we consider three SiD ECal configurations:

\begin{itemize}

\item \textbf{Reduced} (16+8): 16 thin Tungsten layers followed by 8 thick layers.

\item \textbf{Nominal} (20+10): 20 thin Tungsten layers followed by 10 thick layers.

\item \textbf{Ideal} (60+0): 60 thin Tungsten layers and 0 thick, for training a neural network.

\end{itemize}

\noindent In particular, we seek to establish if the reduced ECal performance can approach the nominal performance if the leakage energy of the reduced design is recovered by a neural network trained with the ideal configuration.

The material cost estimate for the nominal SiD ECal was detailed in \cite{Behnke:2013lya}, where high grade ECal Tungsten and Silicon were costed. Excluding labor and any other costs not directly included, the material cost for the reduced, nominal and ideal configurations can be directly calculated from the numbers of thin and thick layers. We estimate the reduced configuration material cost is 21\% lower than the nominal configuration using the TDR costing.


\section{Methodology}

\begin{figure*}
\includegraphics[width=0.5\textwidth]{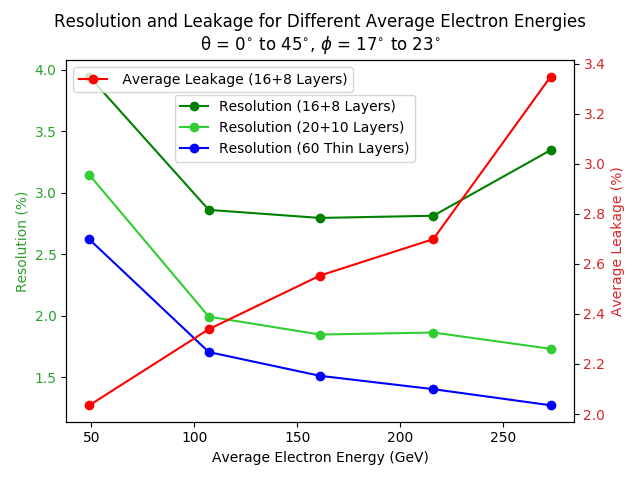}
\includegraphics[width=0.5\textwidth]{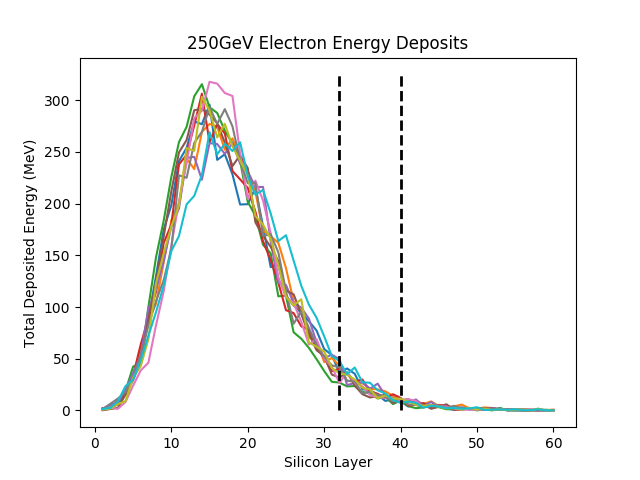}
\caption{At left, energy resolution vs. mean electron energy for the reduced, nominal and ideal SiD ECal outside the overlap regions; also plotted is mean leakage energy for the reduced configuration. At right, energy deposit of 250 GeV electrons vs layer number for the ideal SiD ECal with the simple Geant4 stack. Dashed vertical lines at layer $32=16+2\times 8$ and layer $40=20+2 \times 10$ indicate the energy containment of the reduced and nominal configurations. }
\label{fig:setup}
\end{figure*}

We simulate the ideal ECal with the compact SiD detector description in DD4hep \cite{1742-6596-513-2-022010} using ILCsoft v02-00-02, as well as a standalone Geant4 \cite{Allison:2006ve,ALLISON2016186} simple stack of Silicon and Tungsten slabs for crosschecks. In both cases we generate $10^5$ single electron events with flat angular distributions. The electron energy distribution is flat in the range $20 < E < 300$~GeV. In order to eliminate radiative losses the tracking subsystems are removed and the magnetic field is turned off.

For the energy resolution and mean leakage energy vs. mean electron energy for the reduced, nominal and ideal SiD ECal, see Figure \ref{fig:setup}. See the same Figure for typical shower profiles and leakage with  250 GeV electrons in reduced, nominal and ideal SiD ECal configurations.

We use a multilayer perceptron implemented in TensorFlow \cite{tensorflow2015-whitepaper} to predict the energy leakage measured in the ideal ECal based on the shower profile measured in the nominal and reduced designs. The neural network topology consists of many inputs, a single hidden layer, and one output. The primary inputs are the energy deposited in each ideal ECal layer up to the containment layer, ie layers $1-32$ ($1-40$) for the reduced (nominal) configuration. Additional inputs include the total energy, hit multiplicity, incidence angle \footnote{Throughout, $\theta$ is the angle of incidence of electrons with respect to the ECal face, not the detector polar angle, while $\phi$ is the detector azimuthal angle.} $\theta$ and azimuthal $\phi$. The single output is the predicted leakage energies as measured in the ideal configuration layers $33-60$ ($41-60$) for the reduced (nominal) configuration.

After training the neural network on single electron events, each single electron event energy measurement in the reduced and nominal designs is corrected by adding the energy leakage prediction. The neural network is trained and evaluated using independent event samples. After energy correction the energy resolution of the reduced design is calculated and compared to the energy resolution of the nominal design. 

For further documentation, in greater detail, of the neural network and its performance in the simple Geant4 stack and SiD, the reader is referred to \cite{Braun:2020ivn}.

\section{Results}

\begin{figure}[t]
\includegraphics[width=0.5\textwidth]{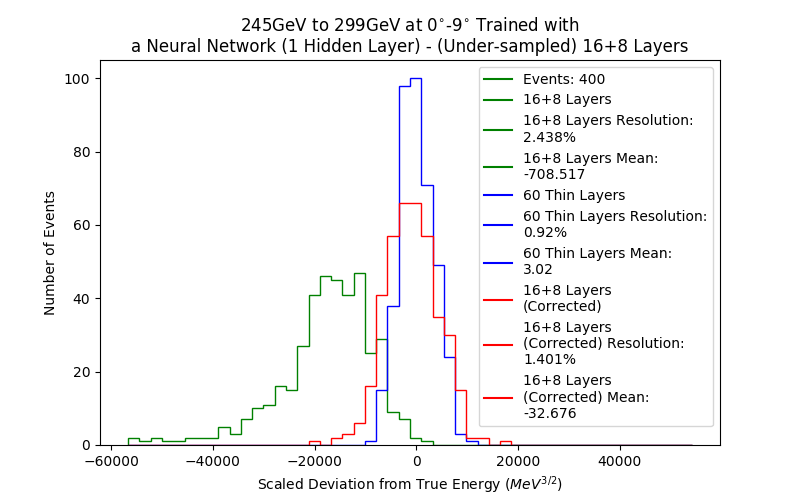}
\caption{Deviation from true energy for reduced (green), reduced corrected (red), and ideal (blue) configurations for electrons with $0 < \theta < 9^{\circ}$ and $245 < E < 299$~GeV in the simple Geant4 stack.}
\label{fig:correction}
\end{figure}

At low angle of incidence and high energy, the leakage energy is maximal and therefore the energy resolution improvement after correction is also maximal. See Figure \ref{fig:correction} for a comparison of energy resolution for reduced, reduced corrected, and ideal configurations with the simple Geant4 stack.

\begin{figure*}[t]
\includegraphics[width=0.5\textwidth]{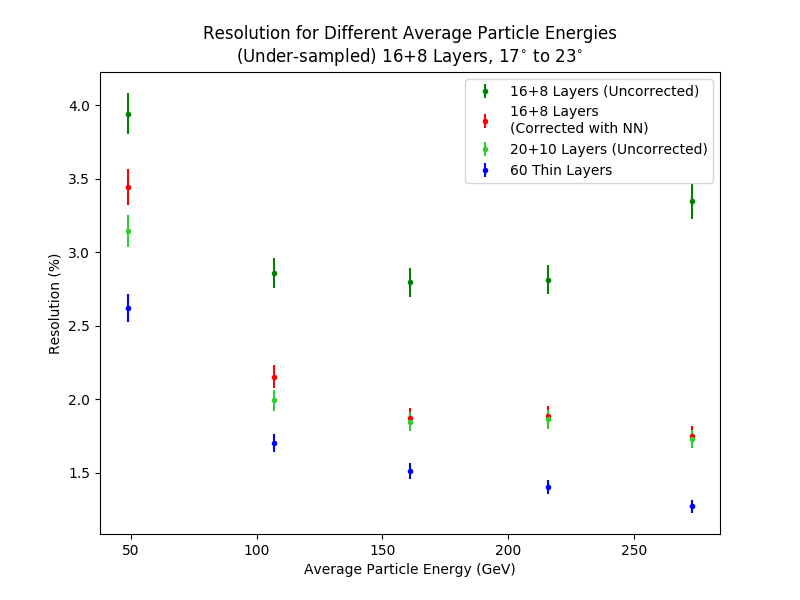}
\includegraphics[width=0.5\textwidth]{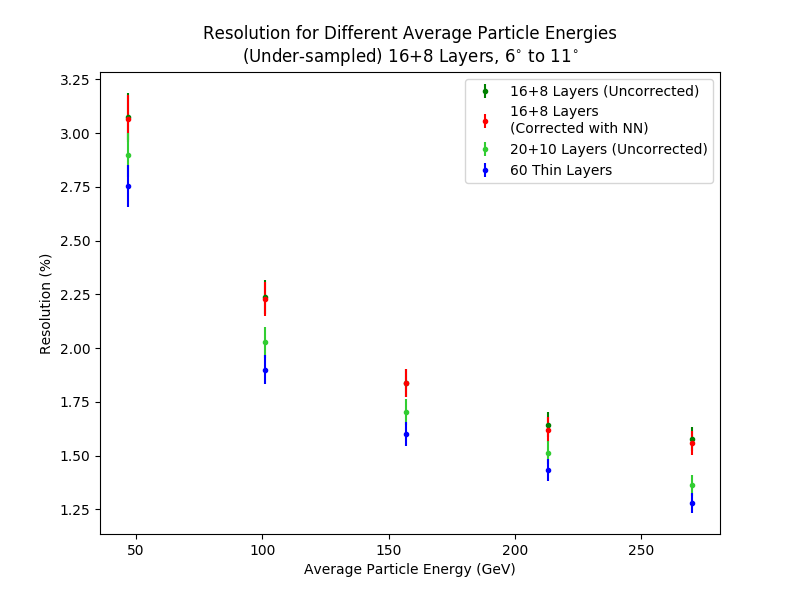}
\caption{Resolution vs mean electron energy for the reduced (dark green), reduced corrected (red), nominal (light green), and ideal (blue) SiD ECal configurations  outside the overlap regions (left) and in the overlap region (right). Simulation employs DD4hep with the compact SiD ECal description in ILCsoft v02-00-02.}
\label{fig:full}
\end{figure*}

For the full SiD ECal simulation with the compact SiD ECal description in ILCsoft, see Figure \ref{fig:full}. Away from the overlap regions, the angle of incidence is low and we sample thick layers, so we expect maximal improvement in energy resolution here. In the overlap regions the angle of incidence is large and all traversed layers are thin, so we expect minimal leakage energy and therefore little improvement in energy resolution, which is already quite good. 

These expectations are confirmed in Figure \ref{fig:full}. In particular, we see that with the leakage correction provided by the neural network, the reduced ECal performance very closely matches the nominal ECal performance. Opting for the reduced ECal configuration may yield a 21\% reduction in material costs with minimal loss in performance.

\section{Conclusion}

The SiD ECal exhibits correlations between Silicon layer energy depositions due to the well understood electromagnetic showering process. Such correlations can be exploited using a neural network to predict energy leakage. Correcting ECal measurements with predicted leakage yields improved energy resolution.

Specifically, the reduced SiD ECal (16+8) performance can nearly match the nominal SiD ECal (20+10) performance by correcting measurements with neural network predicted leakage. Nominal ECal performance can be maintained with fewer layers, and therefore lower cost. Assuming the costing from \cite{Behnke:2013lya}, we estimate a 21\% reduction in material cost results from adopting the reduced ECal design.

We also bring to the reader's attention a similar application of machine learning to calorimeter leakage recovery presented at this conference \cite{masako}.


\bibliography{paper}

\end{document}